# The Young-Laplace equation for a solid-liquid interface

P. Montero de Hijes[1], K. Shi[2], E. G. Noya[3], E. E. Santiso[2], K. E. Gubbins[2], E. Sanz[1], C. Vega[1]
[1] *Faculty of Chemistry, Chemical Physics Department, Universidad Complutense de Madrid, Plaza de las Ciencias, Ciudad Universitaria, Madrid 28040, Spain*
[2] *Department of Chemical and Biomolecular Engineering, North Carolina State University, Raleigh, North Carolina 27606, United States and*
[3] *Instituto de Química Física Rocasolano, Consejo Superior de Investigaciones Científicas, CSIC, Calle Serrano 119, 28006 Madrid, Spain*

The application of the Young-Laplace equation to a solid-liquid interface is considered. Computer simulations show that the pressure inside a solid cluster of hard spheres is smaller than the external pressure of the liquid (both for small and large clusters). That would suggest a negative value for the interfacial free energy. We show that in a Gibbsian description of the thermodynamics of a curved solid-liquid interface in equilibrium, the choice of the thermodynamic (rather than mechanical) pressure is required, as suggested by Tolman for the liquid-gas scenario. With this definition, the interfacial free energy is positive, and the values obtained are in excellent agreement with previous results from nucleation studies. Although for a curved fluid-fluid interface there is no distinction between mechanical and thermal pressures (for a sufficiently large inner phase), in the solid-liquid they do not coincide, as hypothesized by Gibbs.

## I. INTRODUCTION

Under certain conditions (i.e. constant number of particles $N$, volume $V$, and temperature $T$) it is possible to have a spherical phase in equilibrium with another phase around it. There, the Helmholtz free energy $F$ is a local/global minimum representing a metastable/stable equilibrium state [1–8]. This equilibrium implies that $T$ and chemical potential $\mu$ are homogeneous. Thus, $\nabla T(\mathbf{r}) = 0$ and $\nabla \mu(\mathbf{r}) = 0$ where $\mathbf{r}$ is the position vector. However, the number density $\rho(\mathbf{r}) = dN(\mathbf{r})/dV(\mathbf{r})$ and the pressure tensor $\mathbf{p}(\mathbf{r})$ are inhomogeneous [9, 10]. By taking the center of mass of the cluster (COM) as origin and using spherical coordinates,

$$\mathbf{p}(\mathbf{r}) = \mathbf{p_N}(\mathbf{r})[\mathbf{e_r}\mathbf{e_r}] + \mathbf{p_T}(\mathbf{r})[\mathbf{e}_\theta\mathbf{e}_\theta + \mathbf{e}_\phi\mathbf{e}_\phi], \qquad (1)$$

where $\mathbf{e_r}$, $\mathbf{e}_\theta$, and $\mathbf{e}_\phi$ are the unitary vectors. Then, the condition of mechanical equilibrium, $\nabla \cdot \mathbf{p} = 0$, implies [9–11]

$$p_T(r) = p_N(r) + \frac{r}{2}\frac{dp_N}{dr}. \qquad (2)$$

In the late 70's and 80's, Rusanov, Rowlinson, Gubbins and Telo da Gamma and coworkers, pioneered the application of computer simulations to study fluid-fluid spherical interfaces at equilibrium [1, 11, 12]. Lately, there has been a revival in the study of these systems [5, 6, 8, 13–27], including also solid-fluid curved interfaces [4, 7, 28–31]. In fact, we have recently shown that the stable equilibrium observed in the NVT ensemble is an unstable equilibrium in the NpT ensemble that corresponds to a maximum in the Gibbs free energy $G$. Thus, nucleation can be studied via both stable and unstable equilibrium as they are two sides of the same coin [7, 8].

The best thermodynamic description of a system with a curved interface in equilibrium can be found in the book of Rowlinson and Widom [9, 32]. Following in the spirit of Gibbs, one assumes two macroscopic phases that are homogeneous up to the interface and accounts for an additional contribution due to the interface itself. Taking into account that $\mu$ is homogeneous,

$$F = N\mu - p_{int}\frac{4}{3}\pi R^3 - p_{ext}(V - \frac{4}{3}\pi R^3) + 4\pi R^2\gamma, \qquad (3)$$

where $\gamma$ is the interfacial free energy, $R$ the radius of the spherical phase, and $p_{int}$ and $p_{ext}$ the respective internal and external pressures.

At a molecular scale there is some arbitrariness in determining $R$. Since $F$, $\mu$, $p_{int}$, and $p_{ext}$ are fixed, changing $R$ also changes $\gamma$. There are two popular choices for $R$. The first is the Gibbs dividing surface, $R = R_e$, for which the number of excess particles is zero (meaning that particles belong either to the solid or to the liquid, but not to the interfacial region). The second is the surface of tension, $R = R_s$, for which $\gamma$ is a minimum ($\gamma_s$).

Actually, by taking the notational derivative (i.e. an arbitrary change in $R$ without any physical change in the system) one obtains [9, 32]:

$$p_{int} - p_{ext} = \frac{2\gamma}{R} + \left[\frac{d\gamma}{dR}\right]. \qquad (4)$$

By definition, $[d\gamma/dR] = 0$ when $R = R_s$ leading to the celebrated Young-Laplace equation:

$$p_{int} - p_{ext} = \frac{2\gamma_s}{R_s} \qquad (5)$$

Since $\gamma_s$ is positive, this equation shows that the pressure inside the spherical phase is higher than outside and the difference depends on $\gamma_s$ and $R_s$. Simulation studies of fluid-fluid phases by Gubbins and coworkers[11] and Vrabec and coworkers [33] have confirmed the higher pressure of the internal phase. However, this equation has not been tested for a solid-fluid curved interface. This is the goal of this letter.

| Label | $R_s$ | $\rho_{sol}$ | $\rho_{liq}$ | $p_{sol}$ | $p_{liq}$ | $\Delta p$ |
|---|---|---|---|---|---|---|
| IV | 10.791 | 1.0613 | 0.9619 | 12.6046 | 12.7437 | -0.1391 |
| VII | 15.20 | 1.0548 | 0.9560 | 12.3053 | 12.4047 | -0.0994 |
| VIII | 17.467 | 1.0529 | 0.9541 | 12.2199 | 12.3003 | -0.0804 |

TABLE I. Densities and pressures determined at the respective plateaus in the density and pressure profiles. The difference in pressure is also given as $\Delta p = p_{sol} - p_{liq}$. The notation of IV, VII, and VIII refers to the clusters labeled in this way in Ref.[7].

## II. METHODOLOGY

Recently, we simulated several solid clusters in equilibrium with a liquid [7] via the pseudo-hard-sphere PHS continuous potential[34] (hereafter simply HS) which allows to simulate with the standard molecular dynamics package GROMACS[35].

Here, for three selected clusters labeled as IV,VII and VIII in our previous work [7] (see Ref.[7] for further details on the size of these solid clusters and the way they were obtained using NVT simulations), we launch new trajectories (in the NVT ensemble) saving configurations very often allowing us to compute the pressure tensor. Since the definition of the pressure tensor is locally arbitrary, we choose to use the Irving-Kirkwood[36] convention in which the forces between two particles act in the line connecting them. Further detail can be found in the supplementary material (SM). In addition, density profiles are provided.

The simulation details including interaction potential, GROMACS set up, and order parameter to label particles as solid or liquid are exactly the same as in our previous work [7]. We shall use here reduced units. Lengths are given in units of $\sigma$ (i.e the hard sphere diameter), densities as $\rho = N/V\sigma^3$, pressures in units of $(kT/\sigma^3)$, interfacial free energies in units of $kT/\sigma^2$, and chemical potentials in units of $kT$. Pressure profiles are computed up to half of the simulation box $L/2$ whereas density profiles, following Ref.[37], cover the whole system.

## III. RESULTS

The density profile and the normal and tangential components of the pressure tensor are presented in Fig.1. The values of the densities of the solid and the liquid when they reach a plateau, and that of the pressure (far from the interface) are presented in Table I. These are obtained by averaging the data from the corresponding plateaus.

Close to the interface the normal and tangential components of the pressure tensor are different, albeit far from it both are identical. Surprisingly, the pressure inside (solid) is smaller than outside (liquid). This result, in principle, contradicts the Young-Laplace equation. Notice though that having a lower pressure for the solid phase does not violate the mechanical equilibrium condition which only requires a certain relation between $p_N$ and $p_T$ (Eq.2).

This is opposite to the fluid-fluid curved interface. Actually, all previous studies on curved interfaces with fluid phases found higher pressure for the internal phase [1, 11, 12]. Here, for a solid spherical cluster we found lower pressure in the internal phase. One may think that this behavior is peculiar for HS, for which there are no attractive forces. However, recently, Gunawardana and Song[31] have reported a similar behavior for a solid cluster of Lennard-Jones particles surrounded by liquid.

Interestingly, one can already learn the behavior of the curved interface by analyzing the behavior of the pressure tensor for the planar interface. It turns out that in the interfacial region of a planar interface $p_T < p_N$ for fluid-fluid interfaces[38] and $p_T > p_N$ for solid-fluid interfaces [39]. Thus, a simple analysis of the behavior of the pressure tensor for the planar interface is sufficient to know if one will have higher or lower pressure in the internal spherical phase (see Eq.2). It is also interesting to point out that the pressure of the external phase, $p_{ext}$, is identical to the average pressure of the system, $\langle p \rangle$, obtained from the virial equation applied to the entire system provided that the normal and tangential components are identical at $L/2$, as demonstrated in the SM.

Although one must accept the fact that the pressure inside a solid cluster is smaller than the pressure outside, the consequence of this appears to be dramatic as this would imply (apparently) from the Young-Laplace equation that $\gamma$ is negative [40]. The Young-Laplace equation is explained in any textbook of physics, and now we have a problem about how to use it in the case of a solid cluster surrounded by liquid. How to reconcile the results of this work with the Young-Laplace equation? The key was provided by Tolman in his celebrated paper discussing the variation of $\gamma$ with $R$ in a droplet. In particular, there is a remark by Tolman[41] which we believe is highly important in this context. The remark is as follows (adapting his notation to this paper):

*"In applying Eqs. (2.2) and (2.3) to very small droplets, it is to be noted that $p_{int}$ and $\rho_{int}$ are to be taken as the pressure and density for a large mass of internal phase in a condition at the temperature of interest to give the same value of $\mu$ as that of the vapor (cf. Gibbs, reference 1, p. 253)."*

Notice that Tolman was describing the equilibrium between a droplet of liquid in contact with its vapor. However, it also applies for the solid-liquid interface as we are about to show. Therefore, when using the formalism of Gibbs[42], the pressure of the internal spherical phase should not be taken as its actual value, but rather from that of a bulk having the same $\mu$ as the external phase. Similar reasoning was also used by ten Wolde and Frenkel [43].

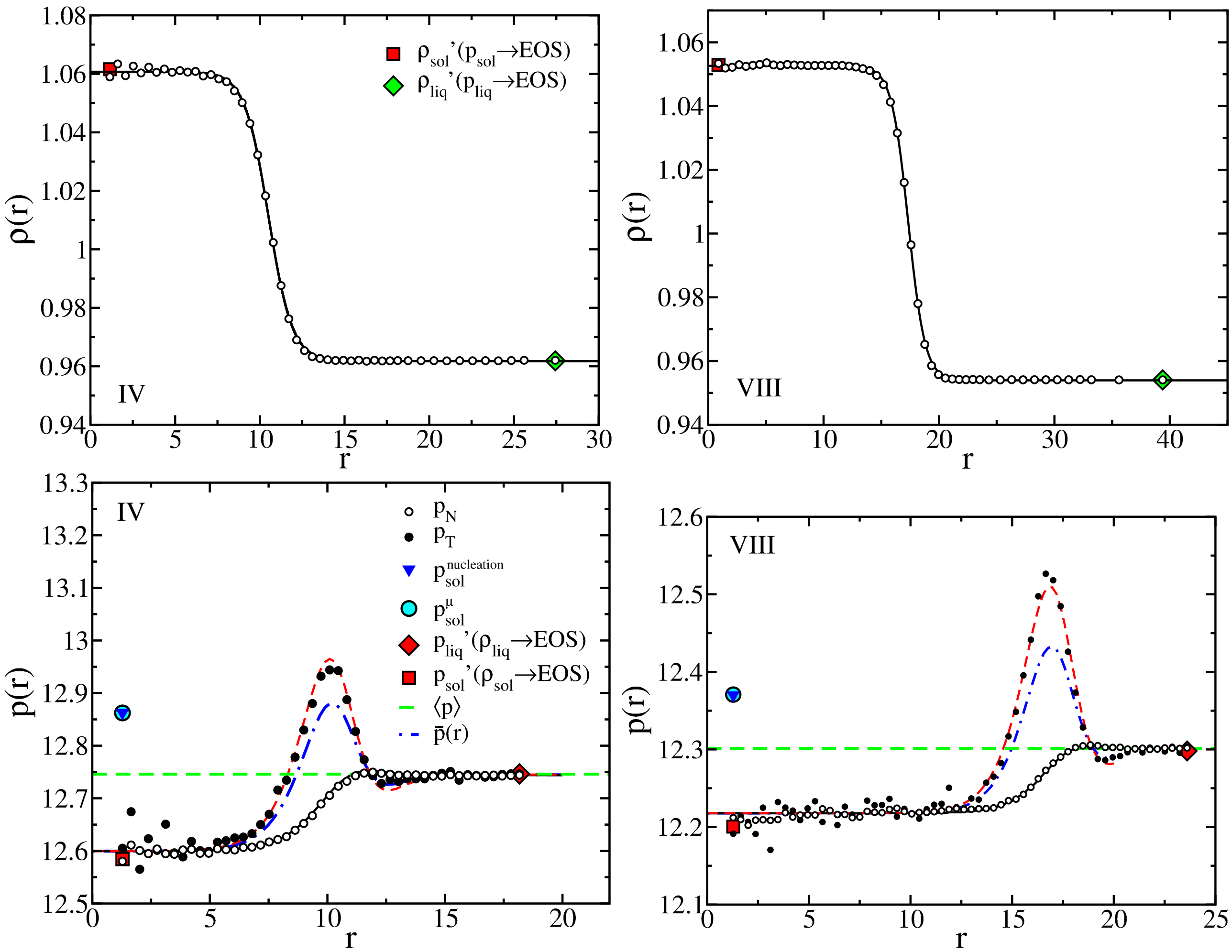

FIG. 1. Radial density (top) and pressure (bottom) profiles from the COM for clusters IV (left) and VIII (right). For the meaning of $p^{\mu}_{sol}$ see the main text. $\langle p \rangle$ is the average pressure of the system as obtained from the virial theorem. $\bar{p}(r)$ is the average pressure at a distance $r$ as given by $(2/3)p_T(r) + (1/3)p_N(r)$. The solid black line is a fit to $p_N$ data, while the red dashed line is obtained from Eq.2 using the $p_N$ fit. In the radial density plot we show the value $\rho'_{sol}$, which would be the density of a bulk solid at $p_{sol}$, and $\rho'_{liq}$, which would be the density of a bulk liquid at $p_{liq}$. In the pressure profile we show $p'_{sol}$, which would be the pressure of a bulk solid having the density $\rho_{sol}$, and $p'_{liq}$, which would be the density of a bulk liquid having the density $\rho_{liq}$. The value labeled as $p^{nucleation}_{sol}$ corresponds to $p^{nucleation}_{sol} = p_{liq} + 2\gamma_s/R_s$ when using the value of $\gamma_s$ and $R_s$ from nucleation studies [7].

Determining the exact value of $\mu$ of inhomogeneous systems of high density is very difficult[44] (notice though some recent progress[45]). Thus, we do not know the exact value of $\mu$ for the three systems considered in this work. However, to illustrate our main point this is not crucial. We shall assume that the external liquid has bulk behavior so that $\mu$ in the system corresponds to that of a bulk liquid at $p_{liq}$. In the inset of Fig.2 a), $\mu(p)$ for solid and liquid bulks are presented (obtained via thermodynamic integration[46] from $p = 11.648$ which is the coexistence pressure[47] where $\mu$ is identical in both phases).

Taking, for instance, system VIII where $p_{liq} = 12.3003$ and $p_{sol} = 12.2199$, it is possible to determine $\mu$ for a bulk liquid phase at this pressure, and also the pressure of a solid that has the same value of $\mu$, which we found to be $p^{\mu}_{sol} = 12.37$. The superscript $\mu$ reminds us that this pressure is not the mechanical pressure of the solid, but rather the pressure of a bulk solid that has the same $\mu$ as that of a bulk liquid at $p_{liq}$. We shall denote this as the thermodynamical pressure (as opposed to the mechanical pressure $p_{sol}$). Notice that $p^{\mu}_{sol} > p_{liq} > p_{sol}$. This finding suggests that the cluster must be different to a perfect bulk at the same pressure $p_{sol}$, otherwise, $\mu$ could not be homogeneous and no equilibrium could be reached. In order to find differences, we followed the evolution of the closest particles to the COM finding that the solid cluster is a "living" structure that can melt in a certain region and grow in another while keeping the size approximately constant. As can be seen in Fig. 2 b), the selected particles ended up quite close to the interface and some of them changed their neighbor. This is likely due to the presence of vacancies in the cluster that lead to a relative diffusion. By computing such diffusion for the cluster as well as for solid bulks with and without vacancies at $p^{VIII}_{sol}$, we could estimate the cluster to have about one vacancy per four thousand particles (1/4000). More relevant is the distribution of distances from a given particle to its 12th closest neighbor considering only particles that fulfill COM $< r < 10\sigma$ in order to avoid surface effects (setting the upper limit in $7\sigma$ did not produce any difference). As shown in Fig. 2 c), such distribution is shifted to higher distances with respect to a bulk with and without vacancies indicating a tiny expansion in the cluster lattice. Further work is needed to completely understand this. Nevertheless, it is clear that the cluster is not identical to a bulk solid at the same mechanical pressure.

| Label | $R_s$ | $p^{\mu}_{sol}$ | $\Delta p^{\mu}$ | $2\gamma_s/R_s$[7] |
|---|---|---|---|---|
| IV | 10.791 | 12.8627 | 0.1190 | 0.1164 |
| VII | 15.20 | 12.4846 | 0.0799 | 0.0793 |
| VIII | 17.467 | 12.3700 | 0.0697 | 0.0694 |

TABLE II. Thermodynamic pressure of the solid and the difference with the pressure of the liquid phase (from Table I). By using the values of $\gamma_s$ from nucleation studies we estimated the term $2\gamma_s/R_s$, and found it to be in excellent agreement with the difference in pressure obtained when using $p^{\mu}$.

Therefore, for the solid-liquid interface the Young-Laplace equation must be written as:

$$p^{\mu}_{sol} - p_{liq} = \frac{2\gamma_s}{R_s} \tag{6}$$

In Table II the value of $p^{\mu}_{sol}$ for the three systems considered in this work is presented along with the difference in pressure $\Delta p^{\mu} = p^{\mu}_{sol} - p_{liq}$.

As can be seen, $\Delta p^{\mu}$ is positive. Thus, by using Tolman's suggestion one recovers a "normal" Young-Laplace equation. In previous work where the same clusters (IV, VII, and VIII) were studied, we obtained the values of $\gamma_s$ from nucleation studies. Since, according to Eq.6, $\Delta p^{\mu}$ corresponds to $2\gamma_s/R_s$, it is of interest to analyze whether our previously reported values of $\gamma_s$ and $R_s$ are consistent with this difference of pressures. As shown in Table II, results are fully consistent. Thus, the physical meaning of $2\gamma_s/R_s$ obtained for the values of $\gamma_s$ and $R_s$ from nucleation studies [7, 48] is now clear. Note that for a fluid-fluid interface there is no difference between $p^{\mu}_{int}$ and $p_{int}$ (for a sufficiently large inner phase) whereas in a solid-liquid system we could not find such agreement even for very large clusters. We have plotted the difference in pressure between solid and liquid as a function of $1/R_s$ in Fig. 2 a). As can be seen, there is no evidence that this difference can become positive for a certain value of $R_s$.

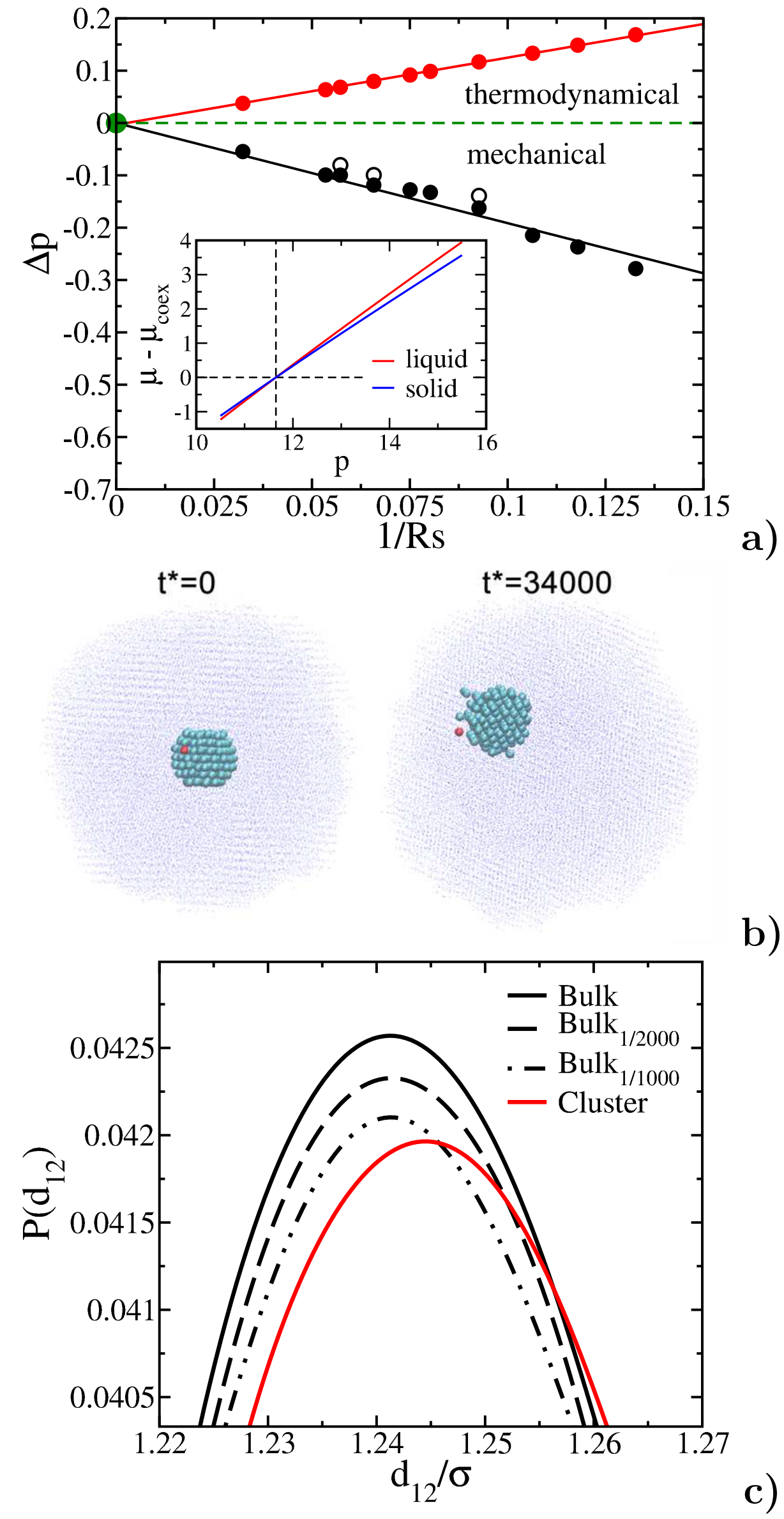

FIG. 2. a) Difference in pressure between the solid cluster and the liquid as a function of $1/R_s$. Upper curve $p^{\mu}_{sol} - p_{liq}$. Lower curve $p_{sol} - p_{liq}$. Note that there are two sets for the mechanical $\Delta p$. Solid black circles are estimated (for all clusters in Ref.[7]) by using the respective densities and the equation of state (EOS), rather than by performing a more costly pressure tensor calculation, as was done for the empty circles. The former is systematically smaller than the latter suggesting that the cluster differs slightly from a bulk. In the inset of this panel, the chemical potential of both phases is shown. b) Snapshots of the solid cluster VIII in the initial configuration and after some time. Only solid particles within the system are shown. We followed the ten closest particles to the COM and their first coordination shells (cyan and red spheres). The remaining solid particles are shown as blue dots. In red, an example solid particle that changed neighbor. c) Maximum in the probability distribution function of the closest twelth neighbor considering cluster VIII and solid bulks with and without vacancies at $p^{VIII}_{sol}$. The ratios 1/1000 and 1/2000 mean the proportion of vacancy per number of particles for the considered bulks. Only the particles at COM $< r < 10\sigma$ were considered.

The idea that for the solid-liquid interface the difference in pressure between the phases may not lead to $\gamma$ was already insinuated by Gibbs. Later, Cahn[49], Cammarata[50, 51], and others [52] suggested that the strain, which is present in solids and not in fluids, was behind this.

## IV. CONCLUSIONS

In summary, we have computed the pressure tensor for a HS system at constant $N$, $V$ and $T$ where one has a stable solid cluster in contact with a liquid away from coexistence conditions. We found that the internal pressure (solid) is lower than the external one (liquid). That would lead to a negative $\gamma$. However, as suggested by Tolman (and insinuated by Gibbs), defining a thermal pressure for inner phase, which corresponds to that of a solid with the same chemical potential as the external liquid, allows to recover a normal Young-Laplace equation, where the pressure of the internal phase is higher, leading to a positive $\gamma$. The values of $\gamma$ from this scheme are in excellent agreement with recent results from nucleation studies. Thus, for a solid-liquid interface one should distinguish between the mechanical and the thermodynamic pressure. This distinction is not so necessary for a fluid-fluid curved interface as they are comparable [25, 51]. However, it is crucial in understanding the meaning of the Young-Laplace equation for a solid-fluid interface. Computer simulations have been the key to solve this subtle issue.

## SUPPLEMENTARY MATERIAL

See the supplementary material for a description of the system and its interaction potential, details on the pressure tensor calculation, values for the fitting parameters as well as the demonstration that the average pressure in the system equals the external pressure. Additional figures and information can also be found.

## ACKNOWLEDGMENTS

This work was funded by grants FIS2016-78117-P and PID2019-105898GB-C21 of the MEC, by project UCM-GR17-910570 from UCM, and by the USA National Science Foundation under Award No. CBET-1855465. PMdH acknowledges financial support from the FPI grant No. BES-2017-080074.

## DATA AVAILABILITY

The data that support the findings of this work are available within the article, and its supplementary material.